\documentclass[aps,twocolumn,showpacs,superscriptaddress,amsmath]{revtex4}
\usepackage{graphicx}
\begin{document}


\title{Torque magnetometry study of metamagnetic transitions in single-crystal
HoNi$_{2}$B$_{2}$C at $\textbf{\textit{T}}\approx$ 1.9 K}
\author{K. D. D. Rathnayaka}
\affiliation{Department of Physics, Texas A\&M University, College
Station, Texas 77843-4242}

\author{B. I. Belevtsev}
\email[]{belevtsev@ilt.kharkov.ua}
\affiliation{B. Verkin Institute for Low Temperature Physics and
Engineering, National Academy of Sciences, pr. Lenina 47, Kharkov
61103, Ukraine}

\author{D. G. Naugle}
\email[]{naugle@physics.tamu.edu}
\affiliation{Department of Physics, Texas A\&M University, College
Station, Texas 77843-4242}



\begin{abstract}
Metamagnetic transitions in single-crystal rare-earth nickel
borocarbide HoNi$_{2}$B$_{2}$C have been studied at $T\approx$ 1.9
K with a Quantum Design torque magnetometer. This compound is
highly anisotropic with a variety of metamagnetic states at low
temperature which includes antiferromagnetic, ferrimagnetic,
non-collinear and ferromagnetic-like (saturated paramagnet)
states. The critical fields of the transitions depend crucially on
the angle $\theta$ between applied field and the easy axis [110].
Measurements of torque along the $c$-axis have been made while
changing the angular direction of the magnetic field (parallel to
basal tetragonal $ab$-planes) and with changing field at fixed
angle over a wide angular range. Two new phase boundaries in the
region of the non-collinear phase have been observed, and the
direction of the magnetization in this phase has been precisely
determined. At low field the antiferromagnetic phase is observed
to be multidomain. In the angular range very close to the hard
axis [100] ($-6^{\circ} \lesssim\phi \lesssim 6^{\circ}$, where
$\phi$ is the angle between field and the hard axis) the magnetic
behavior is found to be ``frustrated''  with a mixture of phases
with different directions of the magnetization.
\end{abstract}


\maketitle

\section{Introduction}
\label{int}
 The rare-earth nickel borocarbides (RNi$_{2}$B$_2$C
where R is a rare-earth element) have attracted considerable
interest in the last decade because of their unique
superconducting and/or magnetic properties (see reviews
\cite{don1,muller,thal,gupta}). The crystal structure of RNi$_{2}$B$_2$C
is a body-centered tetragonal with space group $I4/mmm$
\cite{muller,thal,gupta,sieg,lynn}, a layered structure in which
Ni$_2$B$_2$ layers are separated by R-C planes stacked along the
$c$-axis. The R ions are situated at the corners and in the center
of the crystal unit cell. Conducting (and superconducting)
properties are determined mainly by Ni $3d$ electrons, while
magnetic properties are dictated by localized electrons in the
R $4f$-shell. Long-range magnetic order is thought to result from the
indirect RKKY interaction, mediated through conducting electrons. This gives rise to
different types of antiferromagnetic (AFM) order of the $4f$-ions
at low temperatures \cite{muller,thal,gupta,lynn}. Borocarbides with R =
Tm, Er, Ho, Dy show coexistence of superconductivity and
long-range magnetic order.
\par
Although the borocarbides were studied quite intensively, certain
important issues remain open. We report a torque
magnetometry study of metamagnetic transitions at low temperature
($T\approx 1.9$ K) in a single-crystal magnetic superconductor
HoNi$_{2}$B$_2$C. Torque magnetometry is sensitive only to the
component of the magnetization normal to both the applied
field and the torque and is thus useful in the study of
magnetic anisotropy. The choice of the subject was determined by
its interesting magnetic properties, which cannot be considered as
fully understood to date, and by the circumstance, that torque
magnetometry was applied to the borocarbides so far only in
limited cases \cite{winzer2,canf}. The magnetic properties of
HoNi$_{2}$B$_2$C, a superconductor
with critical temperature T$_c\approx 8.7$~K, are characterized by ({\it i})
large anisotropy and ({\it ii}) availability of different
field-induced magnetic phases at low temperatures
\cite{don1,muller,thal,gupta,lynn,daya,canf,kalats,amici,salamon}.
Three magnetic transitions occur in a narrow temperature interval, when moving from
the paramagnetic state  in low field below $T_c\approx 8.7$ K. The first
two transitions (at 6.0 K and 5.5 K) result in two incommensurate
AFM phases, described in detail in \cite{muller,thal,lynn}.
Below $T_N\approx 5.2$~K, the transition to a commensurate AFM
phase occurs. This AFM phase is a $c$-axis modulated magnetic
structure consisting of Ho-moments ferromagnetically aligned in
the tetragonal basal $(ab)$ planes along the [110] axis and
stacked antiferromagnetically in the $c$-direction. The easy
magnetic axis [110] was found experimentally and supported by
theory \cite{thal,lynn,canf,kalats,amici}. In a tetragonal lattice
four equivalent easy directions $\langle 110 \rangle$ are
expected. The four $\langle 100 \rangle$ axes are hard
directions. No appreciable magnetization was found in the
direction perpendicular to the $ab$-planes (along the $c$-axis) in
fields up to 6~T \cite{canf}, so it is commonly assumed that the Ho
magnetic moments always lie in the $ab$-planes
roughly parallel to one of the $\langle 110 \rangle$ axes.
\par
In the temperature range below $T\approx$ 4 K, several
metamagnetic transitions (of first order) were found, depending on
the magnitude of the magnetic field applied parallel to the
$ab$-planes and the angle $\theta$ between the field and the
nearest easy axis $\langle 110 \rangle$. Below a critical field
$H_{m1}$, the commensurate AFM phase, characterized by
alternating ferromagnetically ordered $ab$-planes with the Ho moments
in neighboring planes in opposite directions, is stable.
This phase can be symbolized by ($\uparrow\downarrow$)
\cite{canf,kalats,amici}, which means that the Ho
moments are parallel to one of the easy directions (say, [110]) in
one half of the $ab$ planes and parallel to the
opposite direction (that is, to {\={1}\={1}0}) in the other half.
\par
With increasing field, at $H=H_{m1}$ a transition to a collinear
ferrimagnetic phase takes place. This phase can be symbolized as
($\uparrow\uparrow\downarrow$). Spins in two
thirds of the $ab$-planes are parallel to one easy axis, while
those in the remaining one third are antiparallel.
It yields one third of the maximum possible magnetization
that the Ho ions can provide. The next metamagnetic transition at
higher field $H=H_{m2}>H_{m1}$ results in the non-collinear phase
($\uparrow\uparrow\rightarrow$) \cite{canf}. It was assumed in the
existing theories \cite{kalats,amici} in this case that two thirds
of the spins are parallel to one of the easy axes (as in
the preceding ferrimagnetic phase), but the remaining one third is
perpendicular to that axis. In this model the net magnetization is thus
predicted to make an angle $\Phi = 26.6^{\circ}$ with the easy axis.
 The nature of this non-collinear phase turns out to be, however, not so simple
\cite{muller,thal,camp,detlef,schneider}. In particular, the models
\cite{kalats,amici} assume a magnetic structure modulated along
the $c$-axis ($q=2/3$~$c^{*}$); whereas, neutron diffraction studies show that
this phase is modulated along the $a(b)$-axis ($q\approx 0.58$~$a^{*}$)
\cite{camp,detlef,schneider}. Finally, at
the field $H_{m3}$, a transition to the ($\uparrow\uparrow$) phase,
the saturated paramagnetic state in which the spins are
aligned ferromagnetically parallel to the $\langle 110 \rangle$
axis nearest to the direction of the applied field, occurs. According to
Ref. \cite{canf}, if the field is directed along or rather close
to a $\langle 110 \rangle$ axis, within the angular region
$-6^{\circ} \lesssim\theta \lesssim 6^{\circ}$, the
($\uparrow\downarrow$)--($\uparrow\uparrow\downarrow$)--($\uparrow\uparrow$)
sequence of transitions takes place. For larger $\theta$ (outside
this range, greater than $6^{\circ}$ and less than
$45^{\circ}$ from the same $\langle 110 \rangle$ axis) the whole
sequence of possible transitions
[($\uparrow\downarrow$)--($\uparrow\uparrow\downarrow$)--($\uparrow\uparrow\rightarrow$)--($\uparrow\uparrow$)]
would be observed. The suggested $H$-$\theta$ diagram of
metamagnetic transitions in HoNi$_{2}$B$_2$C at $T=2$~K based on
longitudinal magnetization measurements \cite{canf} is shown by
the solid lines in Fig.~1.

\begin{figure}[htb]
\includegraphics[width=0.87\linewidth]{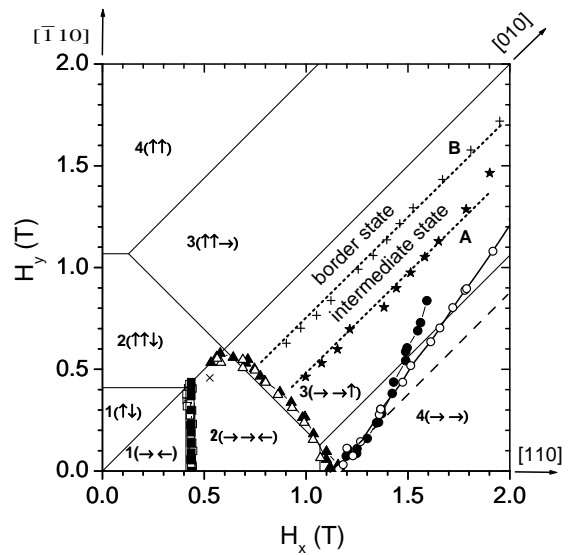}
\caption{The phase diagram for metamagnetic transitions in single
crystal HoNi$_{2}$B$_2$C at low temperature. Symbols of different
magnetic states are defined in the main text of the paper. The
phase-boundary lines (in the angular range between an easy
$\langle 110 \rangle$ and a hard $\langle 010 \rangle$ axis)
indicated by the filled and empty squares, triangles and circles
represent results of measurements at $T\approx 1.9$~K (LM and HM
torque chips, respectively) from this study. The lines A and B,
formed by stars and crosses, are new boundaries of metamagnetic
states revealed in this study. Solid lines represent boundaries
reported \protect\cite{canf} from longitudinal magnetization
measurements at $T\approx 2$~K. The dashed line represents the fit
for data of this study to Eq. (3) with $H_{m3} =0.79$ T at small
$\theta$. The data points $\times$ is indicative of the extent of
frustration in the ($\rightarrow\rightarrow\leftarrow$)
ferrimagnetic state.}
\end{figure}

\par
Since magnetic anisotropy tends to align the Ho magnetic moments
along the $\langle 110 \rangle$ axis, it is clear that angular dependence of the critical
fields ($H_{m1}$, $H_{m2}$, and $H_{m3}$) must be periodic with a
period of $90^{\circ}$ if the four easy axes are equivalent.
The angular dependences of the critical fields in
the $ab$-plane were previously studied mainly with a standard
magnetometer \cite {canf} which measures the magnetization along
the applied field (the longitudinal magnetization). For those studies,
the magnetic field was rotated away from an easy axis [110] in the plane
perpendicular to the $c$-axis. Measurements were taken mainly in the range
$-45^{\circ} \lesssim \theta \lesssim 45^{\circ}$. The results for
the collinear phases [($\uparrow\downarrow$),
($\uparrow\uparrow\downarrow$), ($\uparrow\uparrow$)] were
reasonably well explained by theoretical models
\cite{kalats,amici} which considered the interactions between
moments in an $ab$-plane to be only ferromagnetic. Since neutron
scattering \cite{camp,detlef,schneider} shows that the non-collinear phase
($\uparrow\uparrow\rightarrow$) is, however, not modulated in the
$c$-direction, that phase should be described by a much more
complicated model.
\par
Torque magnetometry has some distinct advantages as compared
with longitudinal magnetometry in studies of magnetically
anisotropic compounds like HoNi$_{2}$B$_2$C. A magnetometer of
this type measures the torque $\vec{\tau}= \vec{M}\times\vec{H}$,
so that $\tau = MH\sin(\beta)$, where $\beta$ is the angle between
external magnetic field and magnetization. This allows precise
determination of the direction of the net magnetization in each phase.
When the magnetization in a material does not align with
the applied magnetic field vector, the torque on the sample is
actually a measure of the magnetic anisotropy energy. The torque
is determined by both, magnitude and direction of the
magnetization. Thus,  changes in magnetization direction (or
rotation of the magnetization) at metamagnetic transitions are
easily seen with torque magnetometry, but the standard
longitudinal magnetometry is much less sensitive, particularly near the
hard axis for this borocarbide.
\par
Three specific problems concerning low-temperature
magnetic states of HoNi$_{2}$B$_2$C have been targeted in this study.
First is the precise determination of the magnetization direction in the
non-collinear phase and its dependence on magnitude of the field. The second
is connected with the four possible equivalent easy direction
$\langle 110 \rangle$ axes. Under these conditions, if a sample is
cooled in zero field, some kind of frustrated or multidomain (or, at least,
two-domain) state can be expected as has been pointed out in studies
\cite{winzer1,winzer2} of the
related magnetic borocarbide DyNi$_{2}$B$_2$C. The third question is
whether different easy $\langle 110 \rangle$ axes are really equivalent
in light of the magnetoelastic tetragonal-to-orthorhombic distortions where
the unit cell is shortened about 0.19 \% along the [110] direction, in which
the long-range ordered Ho moments are aligned \cite{kreysig} as compared to its
length in the perpendicular [$\bar{1}10$] direction at low temperature (1.5 K).
Analysis of the torque behavior in a wide angular range of
magnetic field directions can answer this question.
\par
Results of this study reveal some important new features of the metamagnetic phases
in HoNi$_{2}$B$_2$C and transitions between them, including two new phase boundaries.
Presentation and discussion of results proceed as follows: 1) the angular phase
diagram of metamagnetic transitions reported here
are compared with the only previous detailed study \cite{canf} and with the main
theoretical models \cite{kalats,amici}; 2) some new important peculiarities
of these states and the transitions between them are discussed.

\section{Experimental}
\label{exp}
The PPMS Model 550 Torque Magnetometer (Quantum Design) with
piezoresistive cantilever was used in this study. The HoNi$_{2}$B$_2$C
single crystal  was grown by
Canfield \cite{ming}. The crystal was carefully cut to provide a
sample weighing in the range 0.1--0.15 mg in the form of a plate
($0.4\times 0.32\times 0.26$ mm$^3$) with the $c$-axis perpendicular to
the plate surface and $a$, $b$ axes parallel to the edges. The sample
was mounted on the torque chip in a PPMS rotator so
that the torque measured was along the $c$-axis of the crystal.
During rotation of the sample the $c$-axis was almost precisely perpendicular
to the applied field at all times. This crystal was used previously in our
studies of the magnetic phase diagram of HoNi$_{2}$B$_{2}$C \cite{daya}.
Those measurements were reproduced on several different
single crystals from Canfield's lab. We also made torque measurements
at representative fields and angles on another sample cut from a single crystal from
a different batch of Canfield's crystals and obtained similar results.
Thus, we expect that the results presented here are fully
representative of Canfield's crystals, which seem to set the standard for
work on this family of compounds.
\par
Two types of the PPMS torque chips are used in this study: 1)
Low Moment (LM) (two-leg) chip with RMS torque noise level about
$1\times 10^{-9}$ Nm and maximum allowable torque about
$5\times 10^{-5}$~Nm; and 2) High Moment (HM) (three-leg) chip with RMS torque
noise level about $2\times 10^{-8}$~Nm and maximum allowable torque
about $1\times 10^{-4}$~Nm. The maximal applied fields for LM and
HM chips were limited to 2 T and 3.5 T, respectively, to reduce breakage.
\par
It is known \cite{wille} (but usually not taken into account) that at large
torque the orientation change of the sample due to twisting of the cantilever
itself is no longer negligible. This can cause an error in angular position
of the high-field metamagnetic transitions. The angular coefficient factor for
a silicon piezoresistive cantilever of microscopic size (about 0.2~mm in width)
can be about $(0.3-0.5)^{\circ}/\mu$Nm \cite{wille}. For the HM chip of the PPMS device
this coefficient is about $0.3^{\circ}/\mu$Nm \cite{quantum}. Thus, for higher torque
($\tau >1\times 10^{-5}$~Nm) the angular error for critical fields due to twisting of
the cantilever can be several degrees.
It is also evident that twisting of the cantilever can induce apparent angular
asymmetry of metamagnetic transitions for high magnitudes
of field and torque. We have observed this effect for the critical field $H_{m3}$,
but this asymmetry appears to be rather small (about $1^{\circ}$)
even for the highest fields and torques in this study.
\par
In addition to the error due to twisting of the cantilever, another
important contribution to total error is non-linearity,
when the torque is rather close to the maximum allowable value.
This can cause underestimation of the measured torque (and magnetization) and,
thus, the magnitude of the critical field $H_{m3}(\theta)$. This particular error
should be higher for the LM chip in comparison to the HM chip. In
this study  both the LM and HM chips gave practically the same
numerical  results for the torque and angular dependence of the metamagnetic transitions
below 1.5 T (this field region includes  the first two metamagnetic
transitions for any angular position of the field, and the third transition
to a saturated paramagnetic ($\uparrow\uparrow$) phase for angular positions not
too far from the easy axis).
For the higher field range (1.5--2.0 T), however, the HM chip gave systematically
higher values of the measured torque than the LM chip.
The difference increases with field, so that at $H=2.0$~T the ratio of the torque
values measured by the HM and LM chips is about 1.5. It is clear that above 1.5 T
(where the torque was close to or even larger than $1\times 10^{-5}$~Nm) the HM chip
measures more precisely than the LM one.
\par
In general, we believe that the results of this study obtained below 1.5 T can be
considered as reliable with accuracy for the angle values for magnetic transitions
being about $\pm 1^{\circ}$. For higher field, $H \gtrsim 2$~T, where the torque of
the sample was close to maximum allowable values, the precision is far less. For precise
measurements in higher fields, a sample with smaller mass (and dimensions) should be used,
but we were not able to prepare and orient a significantly smaller sample.
Although the sample studied was mounted carefully with the $c$-axis perpendicular to the
applied field, some misalignment cannot be excluded. It is estimated
that this tilting is no more than $5^{\circ}$, which would produce less than
a 1 \% error in the measured magnitudes of the critical fields of the metamagnetic
transitions. For samples with smaller dimensions this error can be far larger. We have
estimated demagnetization fields to be no more than 30-40 mT at the highest fields which
would affect the accuracy of angular measurements of at most $1-2^{\circ}$.
Torque measurements were made by changing the magnetic field for different
constant angles and by sweeping the angle for different fixed
values of magnetic field at one temperature $T\approx$ 1.9 K only.

\section{Results and discussion}
During discussion of the results obtained, the angular position of
the applied magnetic field will be described by different but related
angles shown in Fig. 2. The first is the angle, $\theta_{r}$, the rotator position
of the torque magnetometer. The second is the angle $\theta$
between the applied field and the nearest easy axis $\langle 110 \rangle$. In
some cases it is more convenient to use the angle $\phi$ between the
field and the nearest hard axis $\langle 100 \rangle$.
Finally, the angle $\beta$ between the external
magnetic field and the magnetization (which does not coincide with
angle $\theta$ for some metamagnetic phases studied) is
also very important for understanding the torque results.

\begin{figure}[htb]
\includegraphics[width=0.9\linewidth]{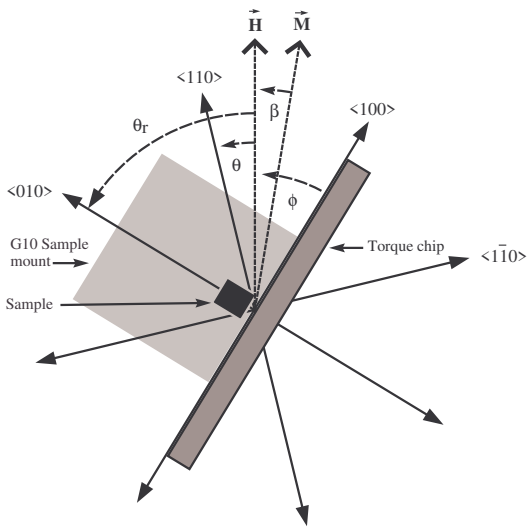}
\caption{Angles used in the discussion. When mounted on the
rotator with $\theta_{r} =0$ the applied field $\vec{H}$ lies
along the $\langle 010 \rangle$ crystal axis.  The sample is
rotated $\theta_{r}$ (indicated by the PPMS rotator) with the
field remaining fixed. The angle $\theta$ is between the nearest
$\langle 110 \rangle$ crystal axis (easy axis) and the applied
field, and $\phi$ is the angle between the nearest $\langle 100
\rangle$ crystal axis (hard axis). The angle between $\vec{M}$ and
$\vec{H}$ is $\beta$.}
\end{figure}

\par
After the sample is mounted in the torque magnetometer, the easy
axes $\langle 110 \rangle$ (or hard axes $\langle 100 \rangle$) in
terms of the angular position of the rotator, $\theta_{r}$, are
initially known only approximately (within a few degrees). But
their location is determined rather precisely from
measured angular dependences of the torque, which are found
to be periodic to a good approximation with a period of
$90^{\circ}$, as expected. The angles $\theta_{r}=45^{\circ}$, $135^{\circ}$ and
$225^{\circ}$ correspond within one degree accuracy to the $\langle 110
\rangle$ easy axes. Similarly, $\theta_{r}=0^{\circ}$, $90^{\circ}$,
and $180^{\circ}$ correspond to $\langle 100 \rangle$ hard axes.

\subsection{Angular phase diagram}
\label{ang}
The angular dependences of the
critical fields for all metamagnetic transitions found in this
study are summarized in Fig. 3 for the whole angular range studied.
They represent results obtained from the
field dependence of the torque recorded for different
angles and angular dependence of the torque recorded for
different fields (see examples in Figs. 4 and 5). The transitions
manifest themselves as sharp changes or even jumps of the torque
at the critical fields or angles. The critical fields
($H_{m1}$, $H_{m2}$, and $H_{m3}$) for a given angle are defined
by the inflection point in the corresponding $\tau
(H)$ transition curves with increasing applied magnetic
field (Fig. 4). Angles of the transitions for different applied fields are
defined in a similar way (Fig. 5). The first
($\uparrow\downarrow$)--($\uparrow\uparrow\downarrow$)
(AFM--ferrimagnetic) transition manifests itself clearly for all
angles, as do other transitions (see Figs. 4 and 5). For angles rather close to the
easy $\langle 110 \rangle$ or hard $\langle 100 \rangle$ axes  some of
the metamagnetic transitions show specific unusual features
outlined further below. The first transition has  considerable hysteresis (Fig.4).
For the other two transitions, the field
hysteresis is small. Some angular hysteresis is also observed when measuring
for increasing and decreasing angle (Fig. 5). For the most part, this hysteresis is
not large (1--1.5 degree) and is isotropic. We believe that
it is mainly due to the backlash in the rotator. For the angular positions
of the field near the hard axis we found an increased angular hysteresis
(up to $4^{\circ}$). For this reason, critical angles of the metamagnetic transitions
recorded for fixed values of the magnetic field were determined for increasing angle.
For some angles the transitions at $H_{m2}$,
($\uparrow\uparrow\downarrow$)--($\uparrow\uparrow\rightarrow$),
and at $H_{m3}$, ($\uparrow\uparrow\rightarrow$)--($\uparrow\uparrow$), show a
change of sign of the torque (Fig. 4), which clearly indicates a change of
direction of the net magnetization from one side of the applied field to the other.
The phase-boundary lines determined in this study are compared (Fig. 1) with the
phase diagram determined from longitudinal magnetization measurements \cite{canf}.
\par
In order to compare the first two transitions with previous measurements, they
are presented in Fig. 6 on an enlarged scale. The first
($\uparrow\downarrow$)--($\uparrow\uparrow\downarrow$)
metamagnetic transition is described well by the angular relation
\begin{equation}
H_{m1}(\theta)=H_{m1}(0)/\cos (\theta)
\end{equation}
[with $H_{m1}(0)=0.437$~T], shown by the solid curve in Fig. 6
(see also the corresponding straight phase-boundary line in Fig.
1). This is consistent with that found from
measurements of the longitudinal magnetization for increasing
field \cite{canf}, except that $H_{m1}(0)\approx 0.41$~T in \cite{canf},
which was fitted with theoretical models \cite{kalats,amici}.
A pronounced hysteresis is found for this transition (Fig. 4) with $H_{m1}$
obtained for decreasing field about 0.04~T below that for increasing field. This
angular dependence (1) is understood in terms of the Ho
moments remaining aligned along the [110] axis when the magnetic
field is rotated from it in the $ab$-plane, as long as the angle of
rotation does not exceed $45^{\circ}$. Thus, only the
projection of the field on the [110] axis is important (see the
phase diagram in Fig.~1).

\begin{figure}[htb]
\includegraphics[width=0.94\linewidth]{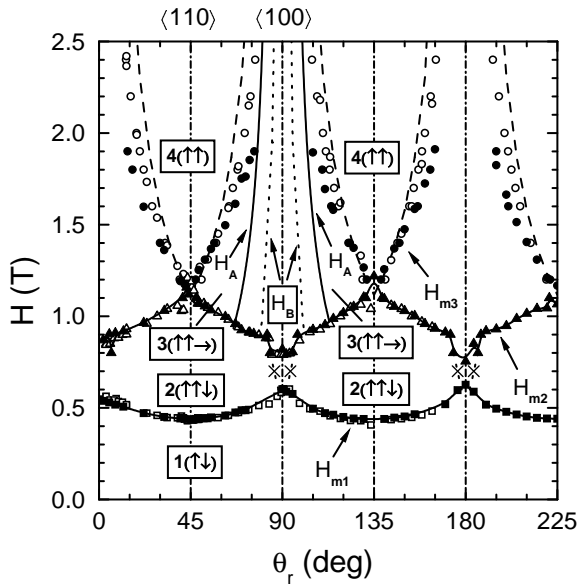}
\caption{The angular phase diagram of metamagnetic states in
HoNi$_{2}$B$_2$C at $T=1.9$~K from torque measurements (increasing
field data). $\theta_{r}$ is the angle on the sample rotator.
$\theta_{r}=45^{\circ}$, $135^{\circ}$ and $225^{\circ}$
correspond to different $\langle 110 \rangle$ easy axes to one
degree accuracy. Symbols for different magnetic states are the
same as in Fig. 1.  $H_{m1}$, $H_{m2}$ and $H_{m3}$ are critical
magnetic fields for transitions between different phases. Easy
$\langle 110 \rangle$ and hard $\langle 100 \rangle$ axes are
marked in one quadrant. Dashed lines in the upper part represent
the theoretical $H_{m3}(\theta)$ dependence (Eq. (3) according to
Refs. \protect\cite{kalats,amici}). Solid lines through data for
$H_{m1}$ and $H_{m2}$ are guides to the eye. Filled and empty
points are obtained with the LM and HM torque chips, respectively.
Data points $\times$ represent the extent of frustration region in
the ferrimagnetic phase estimated from Fig. 11. Solid and dotted
lines ($H_{A}$ and $H_B$) shown in only one quadrant correspond to
new phase boundaries revealed here (the straight lines A and B in
Fig. 1)}.
\end{figure}

\par
For the second metamagnetic transition, the models
\cite{kalats,amici} give the following angular dependence
\begin{equation}
H_{m2}(\phi)=H_{m2}(0)/\cos (\phi),
\end{equation}
where $\phi$ is the angle between the field and the nearest hard
axis $\langle 100 \rangle$ (the line perpendicular to [010] in
Fig. 1 represents this relation). In the previous longitudinal magnetization
study \cite{canf}, good agreement with Eq. (2) was
found with $H_{m2}(0)=0.84$~T.  It also provides a good fit to the data
in Fig. 6 (with $H_{m2}(0)=0.88$~T), but only for angles not too close to the hard
axis $\langle 100 \rangle$, which is the reference axis for this
particular transition (see also the corresponding phase-boundary line
representing this data in Fig. 1). The peculiar behavior of
$H_{m2}$ in the range $-6^{\circ} \lesssim \phi \lesssim
6^{\circ}$ near the hard axis (Fig. 6) will be discussed further below.
This contrasts with Ref. \cite{canf} where good agreement with Eq.~(2) was
found for all angles except near the easy axis.

\begin{figure}
\includegraphics[width=0.90\linewidth]{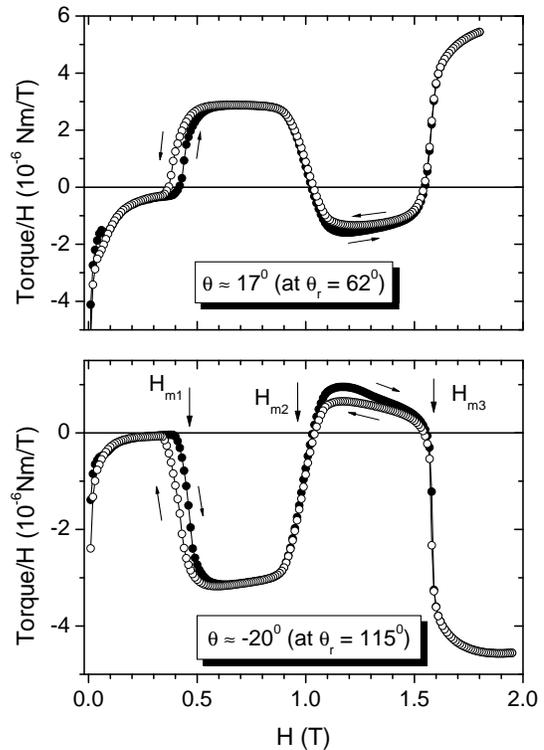}
\caption{$\tau(H)/H$ curves for  angles $\theta \approx
17^{\circ}$ (upper panel) and $\theta \approx -20^{\circ}$ (lower
panel) relative to a closest $\langle 110 \rangle$ axis (rotator
positions in parentheses). Data was recorded for increasing and
decreasing magnetic field (arrows) using the LM torque chip. Three
metamagnetic transitions ($H_{m1}$, $H_{m2}$ and $H_{m3}$) are
marked.}
\end{figure}

\par
It is emphasized that experimental angular dependences of critical fields
for the first two transitions, $H_{m1}(\theta)$ and $H_{m2}(\theta)$, obtained with
the LM and HM torque chips are essentially the same (see Figs. 3 and 6).
A marked difference between the readings of the LM and HM chips appears only for
fields above 1.5 T when the transition to the saturated paramagnetic
(ferromagnetic-like) phase at the critical field $H_{m3}$ takes place
(Figs. 1 and~3).
For this transition the models \cite{kalats,amici} give the expression
\begin{equation}
H_{m3}(\phi)=H_{m3}(0)/\sin (\phi).
\end{equation}
In Ref. \cite{canf} a rather good fit to this expression with
$H_{m3}(0)=0.66$~T was found (the two solid lines parallel to
$\langle 010 \rangle$ in Fig. 1). The dashed lines in the upper
part of Fig. 3 (drawn for $H_{m3}(0)=0.79$~T) and that in Fig. 1 represent a fit of
our data to this equation. It is seen that the agreement
can be considered to be satisfactory only for angular positions
close to the easy axis (low field region). For angular positions
closer to the hard axis, the experimental points deviate more strongly
from the prediction of Eq.~(3). It is seen that for the HM chip this deviation
is much less  than for the LM chip in the range 1.5--2.0~T, but above 2 T data
from the HM chip deviate significantly as well. This is thought to be mainly due to
rotation of the sample and non-linear effects of the cantilever (Sec. \ref{exp})
where angular error can be
several degrees. The deviation from the dashed lines in Fig. 1 and Fig.~3
at higher field is to a large extent, but probably not totally, due to these errors.

\begin{figure}
\includegraphics[width=0.92\linewidth]{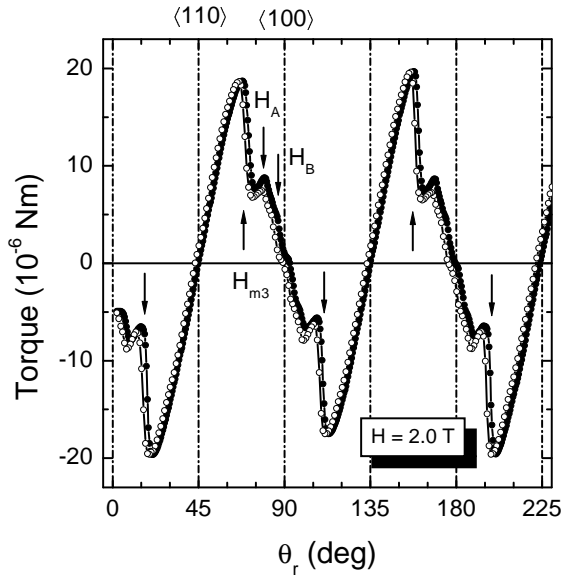}
\caption{Angular dependence of torque at $H=2.0$~T (HM torque
chip). $\theta_{r}$ is the angle on the sample rotator. Easy
$\langle 110 \rangle$ and hard $\langle 100 \rangle$ axes are
shown in one quadrant. Filled and empty circles correspond to data
for increasing and decreasing angle, respectively. Arrows show
angular positions of $H_{m3}$ (transition between
($\uparrow\uparrow\rightarrow$) and ($\uparrow\uparrow$) phases)
and fields $H_{A}$ and $H_B$  (in one quadrant) for two new phase
boundaries (see also Fig.3).}
\end{figure}

\begin{figure}
\includegraphics[width=0.92\linewidth]{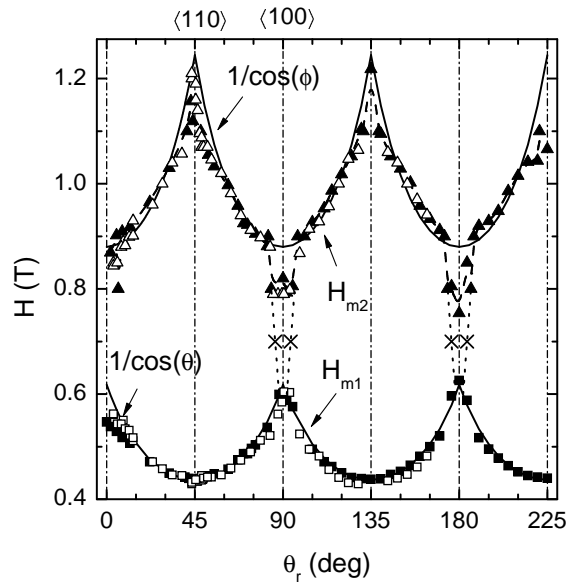}
\caption{Enlarged portion of Fig.~3 presenting the first two
transitions. Solid curve $H_{m1}$ represents
$H_{m1}(\theta)=H_{m1}(0)/\cos (\theta)$ with $H_{m1}(0)=0.437$~T
for the [($\uparrow\downarrow$)--($\uparrow\uparrow\downarrow$)]
transition ($\theta$ is the angle between the field and the
nearest easy axis). Curve $H_{m2}$ represents
$H_{m2}(\theta)=H_{m2}(0)/\cos (\phi)$ with $H_{m2}(0)=0.88$~T for
the
[($\uparrow\uparrow\downarrow$)--($\uparrow\uparrow\rightarrow$)]
transition ($\phi$ is the angle between field and nearest hard
axis). Symbols are the same as in Fig. 3.}
\end{figure}

\par
Some general comments regarding the angular phase diagram for
metamagnetic states in HoNi$_{2}$B$_2$C obtained in this study can
be made. First, the diagram in Fig. 3 appears to be quite periodic
(with a period of $90^{\circ}$). This implies, that low-temperature
orthorhombic distortions of the tetragonal lattice of
HoNi$_{2}$B$_2$C, found in Ref. \cite{kreysig}, do not cause an appreciable
disturbance of angular symmetry
of the metamagnetic transitions. It cannot be ruled out, however, that the
orthorhombic distortions can cause corresponding distortions in
magnetic order of the metamagnetic states \cite{muller,kreysig}.
Possible indications of these effects in the results obtained will be
considered in more detail below in the discussion of particular features of
the metamagnetic states (Sec. \ref{features}). Second, according to
Ref. \cite{canf}, for magnetic field directions close to a $\langle 110
\rangle$ axis ($-6^{\circ} \lesssim\theta \lesssim 6^{\circ}$), solely the
($\uparrow\downarrow$)--($\uparrow\uparrow\downarrow$)--($\uparrow\uparrow$)
sequence with only two transitions ($H_{m1}$, $H_{m2}$)
is observed (see solid lines in Fig. 1).
In this sequence, the transition to the non-collinear
($\uparrow\uparrow\rightarrow$) phase is not present. In contrast,
one model \cite{amici} indicates that this sequence of two
transitions is possible at $\theta=0$ only. Analysis of the
magnetic-field dependences of the torque for different angles,
including angles close to $\theta=0$, indicates that the angular
range for this sequence of only two metamagnetic transitions
($H_{m1}$, $H_{m2}$) is only $|\theta|< 1^\circ$, far less than that
indicated in \cite{canf} (compare the phase boundaries from data of the two studies
in Fig. 1), i.e., the torque measurements for the angles
$|\theta|\geq 1^\circ$ showed three transitions $H_{m1}$, $H_{m2}$ and $H_{m3}$.
 Thus,
torque results support the assertion in \cite{amici} that the sequence of transitions
($\uparrow\downarrow$)--($\uparrow\uparrow\downarrow$)--($\uparrow\uparrow$)
occurs only for $\theta=0$.

\begin{figure}
\includegraphics[width=0.94\linewidth]{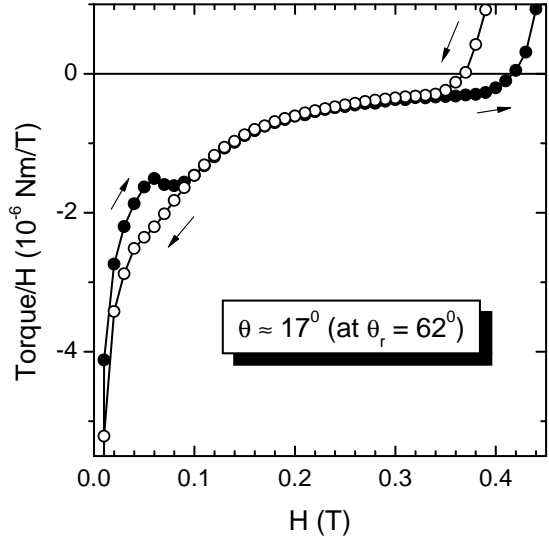}
\caption{$\tau(H)/H$ dependence at $\theta \approx 17^{\circ}$
($\theta_{r}=62^{\circ}$). Filled and empty circles indicate
increasing and decreasing field, respectively.}
\end{figure}

\subsection{Remarkable features of the metamagnetic states}
\label{features}
In this subsection some surprising features of the metamagnetic
states and transitions in HoNi$_{2}$B$_2$C
are considered. First, the net magnetization of the AFM phase must be
equal to zero, and the same should be expected for the torque (magnetic
field less than $H_{m1}$). The experimental evidence is
inconsistent with this. In Fig.~4, the magnetic-field
dependence of the torque (divided by field), $\tau(H)/H=M\sin(\beta)$,
is shown for two angles $\theta_{r}$. A blow-up for one of the angles is
shown in Fig. 7. The $\tau(H)/H$ magnitude is non-zero below $H_{m1}$, but
approaches zero just below $H_{m1}$. Also, $\tau(H)/H$ is hysteretic in the
region $H < 0.1$~T, but above this up to $H_{m1}$ the curves for increasing
and decreasing field coincide.  Such
dependences, with $\tau(H)/H$ being negative below $H_{m1}$, are
found for most of the angular range for angles
$\theta$ of either sign (compare $\tau(H)/H$ curves in Fig. 4 for
$\theta \approx 17^{\circ}$ and $\theta \approx -20^{\circ}$).
Only on the margins of the range investigated ($0 \lesssim \theta_{r}
\lesssim 25^{\circ}$ and $180^{\circ}\lesssim \theta_{r}\lesssim
230^{\circ}$) are the $\tau(H)/H$ values positive below $H_{m1}$
(see upper panel of Fig.~8), but all other features are the same.
Generally, the modulus of $\tau(H)/H$ approaches zero with
increasing field before the first transition starts (Figs. 4, 7
and 8). Opposite signs for $\tau(H)/H$ in different angular ranges
may be an indication of non-equivalence of easy axes $\langle
110 \rangle$ (at least for the AFM state).
\par
The non-zero absolute value of $\tau(H)/H=M\sin(\beta)$ implies
that $M\neq 0$ as well. This may be possible for the AFM state if
a multidomain AFM structure exists. This can be justified by
availability of four (or at least two) equivalent easy $\langle
110 \rangle$ directions in HoNi$_{2}$B$_2$C, as described above.
In such case, on cooling below the N\'{e}el temperature
domains can easily appear \cite{morrish}. The low-temperature orthorhombic
distortions facilitate this process since a shortening of the crystal
lattice can take place along the $\langle 110 \rangle$ directions
in which the moments align \cite{kreysig}. When a
multidomain (or, most likely, two-domain) AFM structure exists, the
magnetic moments of the domains may not be completely compensated,
and the torque may be non-zero. The decrease of torque to nearly zero
with increasing field suggests a transformation to a one-domain state.

\begin{figure}
\includegraphics[width=0.94\linewidth]{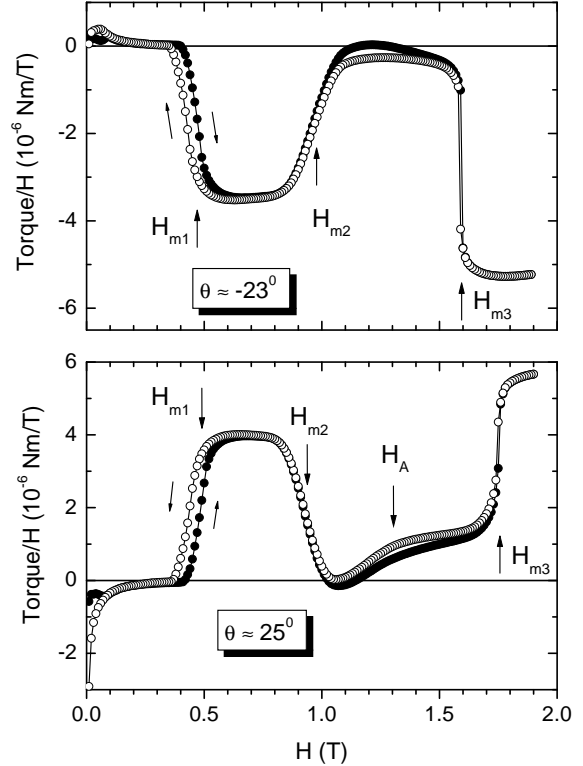}
\caption{The magnetic-field dependence of torque (divided by
field) for  angles close to predicted direction for magnetization
in the $\uparrow\uparrow\rightarrow$ phase  $\theta \approx
-23^{\circ}$ (upper panel) and $\theta \approx 25^{\circ}$ (lower
panel) relative to the closest $\langle 110 \rangle$ axis (rotator
angle in parentheses). Data for increasing and decreasing field
(arrows) was taken with the LM piezoresistive chip. Critical
fields $H_{m1}$, $H_{m2}$, $H_{m3}$  and $H_{A}$ are indicated by
arrows.}
\end{figure}

\par
It should be mentioned that the upper critical field $H_{c2}$ for HoNi$_{2}$B$_2$C
is about 0.3 T at $T=2$~K \cite{daya,lin}. Thus, the torque hysteresis in the AFM
phase is in the superconducting state. Therefore, the non-zero
torque and hysteresis in the low-field range of the AFM state (Fig. 7) may
also be related to trapped flux generated on passing through the
critical field, both as $H$ is increased and decreased. (Note:
data is taken as the field is increased and decreased at a fixed
angle. The angle is then changed to the next value at
approximately zero field, still at 1.9 K.) In  Ref. \cite{lin}, a
noticeable hysteresis in the magnetization at $T=2$~K in polycrystalline
HoNi$_{2}$B$_2$C was found  at $H<0.1$~T, which is consistent with the torque
behavior found in this study.  Recent Bitter decoration experiments \cite{vinnik}
indicate the possibility that pinned vortex structures could be important in
the AFM phase. They suggest that the pinning may arise from AFM twin magnetic
grain boundaries which supports the argument that multidomain
effects explain these low field non-zero torques in the AFM phase.

\begin{figure}
\includegraphics[width=0.94\linewidth]{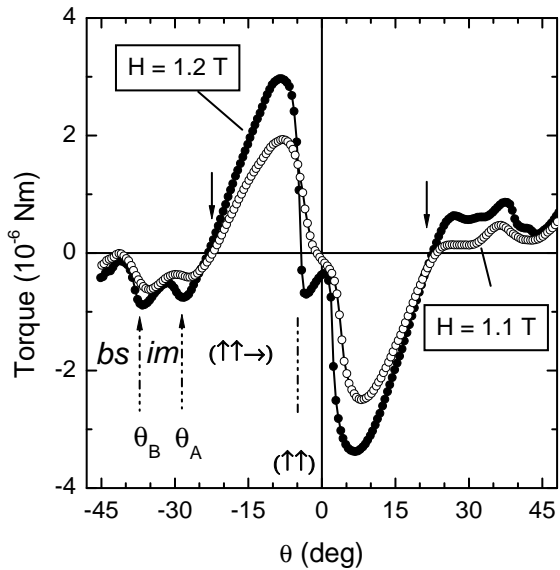}
\caption{Angular dependence (increasing angle) of the torque at
$H=1.1$~T and $H=1.2$~T  with the LM chip.  With increasing
$\theta$ (angle to a closest $\langle 110 \rangle$ axis) torque
changes abruptly for the metamagnetic transition between the
($\uparrow\uparrow\rightarrow$) and ($\uparrow\uparrow$) phases
(near $\theta\approx 3^{\circ}$ and $\theta\approx -3^{\circ}$ for
$H=1.2$~T). Generally, with increasing $\theta$ the sequence of
transitions,
($\uparrow\uparrow$)--($\uparrow\uparrow\rightarrow$)--(intermediate
state)--(border state), occurs (symbols $\it im$ and $\it bs$
represent the last two phases discussed in the main text). Zero
crossing of torque at $|\theta| \approx 23^{\circ}$ (vertical
arrows) indicates the direction of $\vec{M}$ in the non-collinear
($\uparrow\uparrow\rightarrow$) phase.}
\end{figure}

\par
The transition, from the ferrimagnetic
($\uparrow\uparrow\downarrow$) to the non-collinear
($\uparrow\uparrow\rightarrow$) state, at $H=H_{m2}$ is rather
broad, but with negligible hysteresis (Figs. 4 and 8). According
to the model \cite{kalats}, the magnetization in the phase
($\uparrow\uparrow\rightarrow$) is tilted by an
angle $\Phi = \arctan (1/2) \approx 26.6^{\circ}$ to the easy
$\langle 110 \rangle$ axis closest to the magnetic field, and its
absolute value is equal to $\approx 0.745$ of the easy-axis
saturation value. Thus, not only does a change in magnetization
magnitude take place at this transition, but also the angle between the
magnetization and applied field changes. For the
torque $\vec{\tau}=\vec{M}\times\vec{H}$ with the modulus $\tau =
MH\sin(\beta)$, this implies that the angle $\beta$ will be
changed $26.6^{\circ}$ at this transition. The angles $\beta$ and
$\theta$ can be taken as identical in the ferrimagnetic phase. Thus relation $\beta =
\theta - \Phi$ should hold after transition to the non-collinear
phase. This is verified very well in this torque study. A change
in sign of the torque at the transition is expected for field
directions, for which $|\theta| < \Phi$ and is found (Fig. 4). Also the torque
must be close to zero after the transition for $\theta\approx\Phi$. This
specific case is also confirmed (Fig. 8). Since the
torque changes sign as $\beta$ changes sign when
$\theta = \Phi$ in the field range for existence of the
($\uparrow\uparrow\rightarrow$) phase, $\Phi$ can be
precisely determined from the angular dependence of the torque.
Figure 9 presents such dependences for fields 1.1 T and 1.2 T. For both fields
a change in sign of the torque occurs at $\theta \approx 23^{\circ}$, which is
close to but smaller than the theoretical value of $\Phi$.
For fields $H=1.4$~T and 1.5 T this angle is about $22^{\circ}$ and $20^{\circ}$,
respectively, suggesting that $\Phi$ depends on the applied field.
\par
As mentioned above, it is possible, using the angles $\beta$ for each
phase as described above,  to calculate the magnetization magnitude
for different metamagnetic states of the sample by dividing the torque
measured in the range of each phase by $H \sin(\beta)$.
The results are:
1) $M_{\uparrow\uparrow\downarrow} = (0.9-0.95) \times 10^{-2}$~emu
(for the ferrimagnetic phase);
2) $M_{\uparrow\uparrow\rightarrow}\simeq 1.8 \times 10^{-2}$~emu
(for the non-collinear phase at $H=1.6$~T);
3) $M_{\uparrow\uparrow}\simeq 2.8 \times 10^{-2}$~emu
(for the ferromagnetic-like phase at $H=1.6$~T).
It is seen that $M_{\uparrow\uparrow}/M_{\uparrow\uparrow\downarrow} \approx 3$ in
line with the model \cite{kalats}.  The ratio
$M_{\uparrow\uparrow\rightarrow}/M_{\uparrow\uparrow}$ is about 0.64 according to our
estimates. This is somewhat less than that (0.745) predicted by the model
\cite{kalats}. It should be noted, however, that 0.745 is
expected only together with $\Phi=26.6^{\circ}$. For a smaller $\Phi$,
the magnetization of the non-collinear phase should be smaller, so that a
somewhat diminished ratio $M_{\uparrow\uparrow\rightarrow}/M_{\uparrow\uparrow}$ is
quite expected. Furthermore, there is no reason to expect that the
simple model \cite{kalats,amici} which assumes only
 ferromagnetic coupling of the $ab$-planes would predict
$\Phi$ (or $M_{\uparrow\uparrow\rightarrow}$) precisely when the
modulation vector measured by neutron scattering \cite{camp,detlef,schneider}
is $q\approx 0.58$~$a^{*}$, not $q=2/3$~$c^{*}$ as predicted for the
($\uparrow\uparrow\rightarrow$) phase \cite{kalats,amici}.

\begin{figure}[h]
\includegraphics[width=0.92\linewidth]{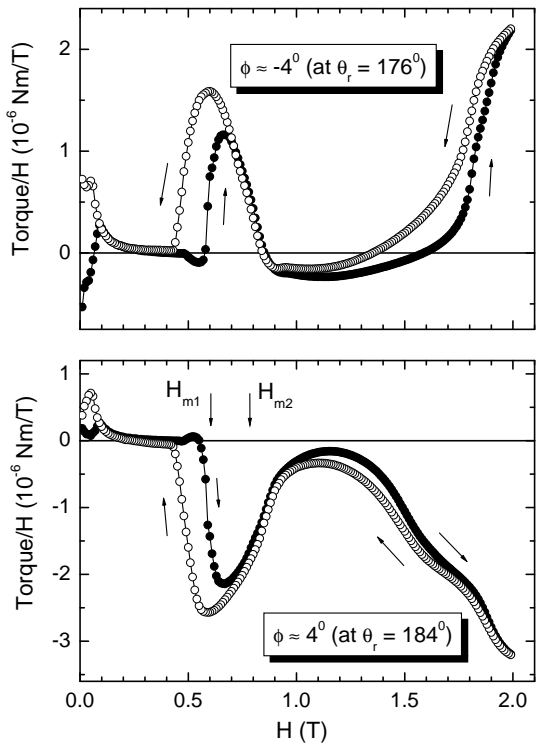}
\caption{Dependences $\tau(H)/H$ for increasing and decreasing
field (arrows) for  angles $\phi \approx -4^{\circ}$ (upper panel)
and $\phi \approx 4^{\circ}$ (lower panel) relative to the closest
$\langle 100 \rangle$ hard axis (rotator angles in parentheses).
The metamagnetic transitions
($\uparrow\downarrow$)--($\uparrow\uparrow\downarrow$) and
($\uparrow\uparrow\downarrow$)--($\uparrow\uparrow\rightarrow$) at
$H_{m1}$ and $H_{m2}$ for angles $\phi$ near 0 are quite close
together, indicative of frustrated behavior.}
\end{figure}

\par
As mentioned earlier, the angular behavior of $H_{m2}$ for angles
very close to the hard axis $\langle 100 \rangle$ ($-6^{\circ}
\lesssim \phi \lesssim 6^{\circ}$) deduced from the torque
measurements is not described by Eq.~(2). The $H_{m2}$ values
in this region are far below those predicted by Eq.~(2) (Figs. 3
and 6). Behavior of $\tau(H)/H$ for small values of $\phi$ is shown in
Fig. 10 to help understand this phenomenon. It can be seen
that the magnetic-field behavior of the torque in the AFM
state is similar to that for angles $\phi$ far from zero (compare
Fig. 10 with Figs. 4, 7, and 8). The first transition to the
ferrimagnetic state manifests itself quite clearly,
although with larger hysteresis. With a further increase in
field the system behaves rather peculiarly: the first
($\uparrow\downarrow$)--($\uparrow\uparrow\downarrow$) and the second
($\uparrow\uparrow\downarrow$)--($\uparrow\uparrow\rightarrow$)
metamagnetic transitions are much closer together for small $\phi$ than expected.
The phase boundaries may not really be changed, but the system cannot
decide which direction its magnetization should point, to the left or
to the right of the hard axis. This behavior
of the magnetic system studied for small $\phi$ is characteristic of ``frustration''.

\begin{figure}
\includegraphics[width=0.92\linewidth]{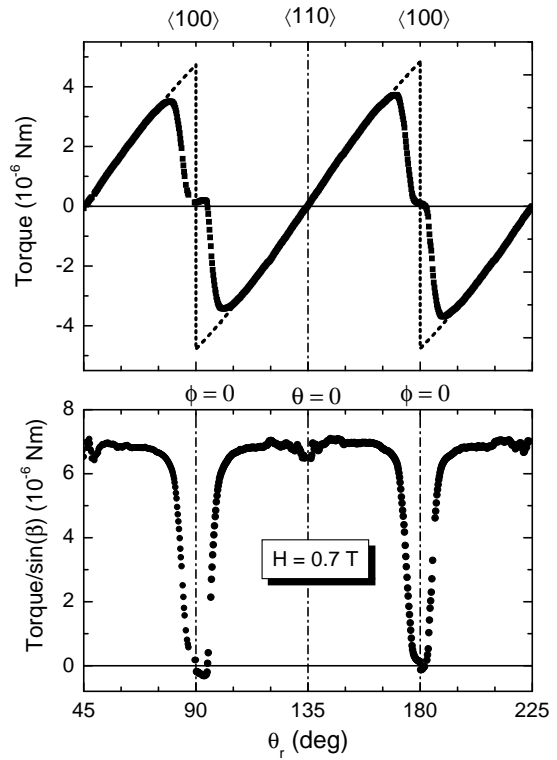}
\caption{Angular dependence of  torque for increasing angle with
the LM piezoresistive chip (upper panel) and $\tau/\sin(\beta)$
(lower panel) at $H=0.7$~T where only the ferrimagnetic
($\uparrow\uparrow\downarrow$) phase exists and $\beta=\theta$
($\theta_{r}$ is angular position of the rotator). Easy $\langle
110 \rangle$ and hard $\langle 100 \rangle$ axes with the
corresponding values $\theta=0$ and $\phi=0$, are shown. The
dashed line indicates the expected dependence of torque near
$\phi=0$ upon the change from one ferrimagnetic state
($\uparrow\uparrow\downarrow$) to another
($\rightarrow\rightarrow\leftarrow$). The extent of the frustrated
region in the ferrimagnetic state is indicated by the region of
almost zero torque near $\phi =0$.}
\end{figure}

\par
Additional  important features of ``frustration'' near
$\phi=0$ were found in the angular dependences of the torque.
Figures 5, 11 and 12 reveal that the torque goes to zero as the direction of the
magnetic field approaches $\phi=0$ from any side in the angular region of
the ferrimagnetic ($\uparrow\uparrow\downarrow$) and non-collinear
($\uparrow\uparrow\rightarrow$) phases (see the phase diagram
in Fig.~3). This behavior of the torque is in sharp contrast to the simple picture
expected from the models \cite{kalats,amici} where the component of the  magnetization
normal to the hard axis changes sign as the field direction crosses  $\phi=0$.
Therefore, $\sin (\beta)$ is expected to change sign as well on crossing the angle $\phi=0$,
producing a very sharp jump in the torque at $\phi=0$ as
depicted by dashed lines in Figs. 11 and 12. The experimental picture is far from that
(compare the experimental and calculated curves in these figures).
\par
It is possible, in principle, to derive angular dependences of the magnetization
(including the ``frustrated'' region near $\phi=0$) from the corresponding angular
dependences of the torque, if the angle $\beta$ is known for all $\theta$.
In this case the values of $\tau/\sin(\beta)$, which should be proportional to
the magnetization, can be  calculated. The angular dependence of
$\tau/\sin(\beta)$ for $H=0.7$~T, calculated with the assumption
$\beta=\theta$, is shown in Fig. 11. It is seen that $\tau/\sin(\beta)$ goes
to zero when approaching $\phi=0$. This takes place approximately
in the range $-6^{\circ} \lesssim \phi \lesssim 6^{\circ}$. Outside these
regions, $\tau/\sin(\beta)$ is approximately constant as expected
for the ($\uparrow\uparrow\downarrow$) phase.
\par
Thus, for the ferrimagnetic phase ``frustration'' proceeds in a rather narrow
angular range $-6^{\circ} \lesssim \phi \lesssim 6^{\circ}$.
This range is estimated by the dotted lines through the data points $\times$ in Fig. 6.
It can be suggested that, for the ferrimagnetic ($\uparrow\uparrow\downarrow$)
phase near the hard axis, a border state develops where $\tau \rightarrow 0$ as
$\phi \rightarrow 0$. When the angular position of the field crosses the
hard axis, ordered moments change their alignment from one easy axis to another.
So that this transition should be accompanied by considerable magnetoelastic
deformation or a change in direction of the orthorhombic distortion, therefore,
being   actually first order. The increased angular hysteresis in the vicinity
of a hard axis,  mentioned above in Sec. \ref{ang}, supports this view.
\par
Although Fig. 11 could be interpreted as $M \rightarrow 0$ as $\phi \rightarrow 0$,
that is inconsistent with longitudinal magnetization measurements
\cite{daya,schneider}. It is more likely that $\sin(\beta)$ goes to zero as the field
direction crosses the angle $\phi=0$. In such an event, the magnetization at $\phi=0$
can be non-zero, but the average magnetization of the sample should be directed along
the hard axis $\langle 100 \rangle$ with $\beta=0$.

\begin{figure}[tb]
\includegraphics[width=0.79\linewidth]{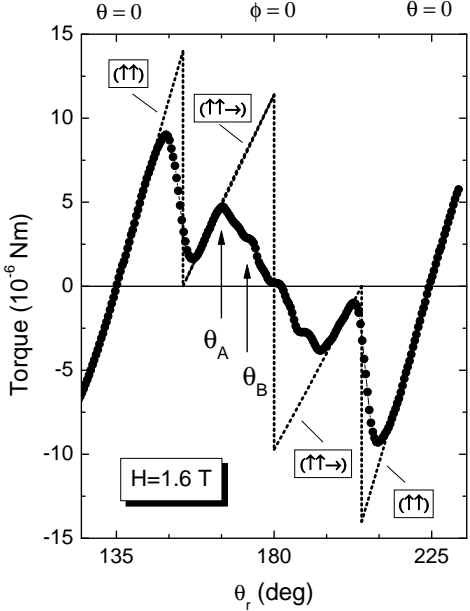}
\caption{Angular dependence of torque at $H=1.6$~T [HM chip,
$\theta_{r}$ rotator angle, easy $\langle 110 \rangle$ ($\theta
=0$) and hard $\langle 100 \rangle$ ($\phi =0$) axes]. Filled and
empty circles represent increasing and decreasing angle,
respectively. Dashed lines show the expected angular dependence of
the torque in the angular areas of the ferromagnetic-like
($\uparrow\uparrow$) and non-collinear
($\uparrow\uparrow\rightarrow$) phases with $\beta=\theta$ for the
($\uparrow\uparrow$) phase and $\beta=\theta-\Phi$ for the
($\uparrow\uparrow\rightarrow$) phase over the predicted range of
these phases ($\beta$ is the angle between the field and the
magnetization, $\Phi$ is the predicted angle \protect\cite{kalats,
amici} between  magnetization and an easy axis in the
$\uparrow\uparrow\rightarrow$ phase). $\theta_{A}$ and
$\theta_{B}$ indicate positions of drastic changes in the torque
corresponding to two new phase boundaries represented by the lines
A and B in Fig.~1. Note increased angular hysteresis near the hard
axis in the region of these two phases.}
\end{figure}

\par
Frustration takes place in two steps in the non-collinear phase.
When $\theta$ increases above some specific angle $\theta_{A}$ (defined as a position
of the maximum in $\tau(\theta)$ in the region of the
($\uparrow\uparrow\rightarrow$) phase in Fig. 12) the torque
decreases rather steeply to a plateau. At somewhat larger angle it decreases rapidly
again at $\theta_{B}$ to a second plateau near zero torque. It is significant that
deviation of torque behavior from
that expected (in line with models \cite{kalats, amici}) for the non-collinear phase
begins well away from $\phi=0$ (Fig. 12). So it appears that, in some substantial
angular range around the hard axis, the metamagnetic state of the sample is different
from the non-collinear phase described in models \cite{kalats,amici}.
Due to the radical change in angular behavior of the torque at the angle $\theta_{A}$
we can consider this point as an indication of a new metamagnetic transition from
the non-collinear ($\uparrow\uparrow\rightarrow$) phase to a different phase which
we will call the intermediate state. The critical angle $\theta_{A}$ for this
transition is found to be strongly field dependent ranging, when measured from a hard axis,
from about $24^{\circ}$ at $H=1.0$~T to about  $4^{\circ}$ at $H=3.2$~T. The corresponding
phase boundary is
represented by a straight line A in Fig. 1 which is parallel to the dashed straight line
of the ($\uparrow\uparrow\rightarrow$)-($\uparrow\uparrow$) transition plotted
according to  models \cite{kalats,amici}. Therefore, the critical
field $H_{A}$, defined by line A in Fig. 1, has the same $1/sin(\phi)$
angular dependence as that given by Eq. (3). The field $H_{A}(\phi=0)\approx 0.35$~T
is roughly one-half of the field $H_{m3}(0)\approx 0.79$~T. The solid line
corresponding to this boundary  $H_{A}$ is shown also in the angular phase
diagram in Fig. 3.
\par
The second specific angle $\theta_{B}$ shown in Fig. 12 is strongly field dependent
as well, defining another transition to a new metamagnetic state which will be called
the border state. The transition indicated by the angle $\theta_{B}$ is represented
by the line B in Fig.~1 and the corresponding critical field $H_{B}$ is also found to
be proportional to $1/\sin(\phi)$ as in Eq. (3) with $H_{B}(0) \approx 0.17$~T, almost
exactly one-half of $H_{A}(0)$.
It can be suggested, therefore, that Figure 12 shows the following sequence of
transitions with increasing $\theta$ from 0 to $45^{\circ}$:
($\uparrow\uparrow$)--($\uparrow\uparrow\rightarrow$)--(intermediate state)--(border state).
According to the phase diagram in Fig.~1, this sequence of transitions should
take place for any field above $\simeq 1.1$ T (but, of course,
with other critical angles). Indeed, $\theta_{A}$ and $\theta_{B}$ can be seen
in Figs.~5 and 9, too.
\par
The occurrence of the intermediate phase was found at first by means of examination
of angular dependences of torque like those shown in Figs. 5, 9 and 12.
The resulting phase diagram (Fig. 1)
implies, however, that this phase should manifest itself in experimental
$\tau (H)$ curves as well for some rather narrow angular range which was estimated to be
$24^{\circ} \leq \theta \leq 32^{\circ}$. Within this range the following sequence of
metamagnetic transitions is expected with increasing field:
($\uparrow\downarrow$)--($\uparrow\uparrow\downarrow$)--(intermediate phase)--($\uparrow\uparrow\rightarrow$)--($\uparrow\uparrow$).
For angles $\theta$ smaller than $\approx 24^{\circ}$ only three
transitions should occur:
($\uparrow\downarrow$)--($\uparrow\uparrow\downarrow$)--($\uparrow\uparrow\rightarrow$)--($\uparrow\uparrow$).
Both these cases are apparent in Fig.~8. In the upper panel, $\tau(H)/H$ is presented
for angle $\theta\approx -23^{\circ}$ (which is equivalent to $\theta\approx 23^{\circ}$
due to the angular symmetry). The magnitude of $\tau(H)/H$ at a
constant angle $\theta$ is equal to the normal component of the magnetization. It is seen
that when transition with critical field $H=H_{m2}$ is completed for high enough field,
the magnetization is nearly constant for  $\theta \approx -23^{\circ}$;
whereas, for $\theta \approx 25^{\circ}$, the magnetization
first increases with increasing field and only for a higher field the magnetization appears
to come to a saturated value. This is where the transition between the intermediate and
the non-collinear phases is expected according to the line A in Fig. 1. This transition
manifests itself more clearly for higher angles $\theta$ (not shown).
\par
We have defined the transition lines A and B in Fig. 1 based on distinct features
in the torque {\it vs.} $\theta$ curves at constant field, but there is uncertainty in the
exact part of the feature that represents the transition. The experimental angular
uncertainty is also more important in this small angular range. Nevertheless, the
$1/\sin(\phi)$ dependence for $H_{A,B}$ in Figs. 1 and 3 is quite precise, though the
``actual'' transition could be shifted parallel to the hard axis slightly in Fig. 1.
Based on the arguments used by Canfield et al. \cite{canf}, the $1/\sin(\phi)$ dependence
implies that the change in magnetization $\Delta \vec M$ at these boundaries
(lines A and B) is normal to the nearest hard axis from the field direction. Thus
$|\Delta \vec M|=\Delta\tau\tan(\phi)/H_{A}(0)$ where $\Delta\tau$ is the change in
torque at the boundary A. With $\Delta\tau$ from torque {\it vs.} $\theta_{r}$ data at fixed
$H=1.3$~T and 1.4~T (not shown, but like that shown in Fig. 9 for $H=1.1$ and 1.2 T and
Fig. 12 for $H=1.6$~T), we estimate $|\Delta M|\approx 4\times 10^{-3}$ emu. Under
the assumption that $M_{\uparrow\uparrow} \approx 2.8\times 10^{-2}$~emu corresponds
to 9.8 $\mu_{B}$ per Ho ion, this $|\Delta M|$ corresponds very roughly to flipping one
Ho$^{3+}$ moment
in 6 of those aligned along the nearest easy axis to the applied field $\vec H$ to the
perpendicular easy axis in going from the non-collinear phase
($\uparrow\uparrow\rightarrow$) to the intermediate state. Although the error
is much larger since the B line can be intercepted only with $\vec H$ less than
$10^{\circ}$ from the hard axis, a similar estimate indicates very roughly that
an additional 2 in 9 Ho$^{3+}$ moments originally along the dominant easy axis
are flipped $90^{\circ}$ in going from the intermediate state to the border state.
\par
Can the available neutron scattering data for HoNi$_{2}$B$_2$C \cite{detlef,camp,schneider}
clarify the new results of this torque study,
concerning new phase boundaries revealed and ``frustration'' behavior near a hard axis?
Data by Detlefs, et al. \cite{detlef} at $15^{\circ}$ to the easy
axis shows the $q\approx 0.58$~$a^{*}$ modulation vector in the region of the
($\uparrow\uparrow\rightarrow$) phase in Fig.~3 while that by Campbell, et al.
\cite{camp} along the hard axis in the region of the border phase shows a
$q\approx 0.61$~$a^{*}$ modulation vector. Measurements by Schneider \cite{schneider}
with increasing field along the hard axis (in the $b^{*}$-direction) at 2 K indicate
the presence of a $0.58$~$a^{*}$ propagation vector in the ferrimagnetic phase which
switches over to a 0.62~$b^{*}$ propagation vector (parallel to the applied field) in
coexistence with a weaker 0.60 $b^{*}$ modulation in the region of our ``border'' phase.
He suggests, based on the argument \cite{detlef} that the $a^{*}$, $b^{*}$
modulations may only develop from non-collinear phases, that the 0.58~$a^{*}$ modulation in the ferrimagnetic
phase region may be associated with the C6 non-collinear phase
 $(\uparrow\downarrow\uparrow\rightarrow\leftarrow\rightarrow)$ predicted by Amici and
Thalmeier \cite{amici}. We note that this phase with
magnetization along the hard axis might possibly develop in the very narrow frustration region
near the hard axis in the ferrimagnetic phase (see Fig. 11), but we did not see any
evidence of a phase boundary in the entire ferrimagnetic region corresponding to that
predicted for C6 \cite{amici}, and the torque measurements outside of the
frustration region were completely consistent with the ($\uparrow\uparrow\rightarrow$)
phase, as were longitudinal magnetization measurements \cite{daya,schneider}. We
note that the boundary between our ``intermediate'' phase and the non-collinear phase
is in the general vicinity of the F2 phase ($\uparrow\rightarrow$) predicted by Amici
and Thalmeier \cite{amici}, but the field-angle dependence is quite different.
We can only speculate that the $b^{*}$-modulations observed by Schneider may be
related to the ``border'' and ``intermediate'' phases observed here.
\par
It is also possible that the ``frustration'' behavior results from a two-domain state
when two possible directions of orthorhombic distortions are realized simultaneously
when the field is directed close enough to the hard axis (so that there is some mixed
state around the hard axis rather than a single phase). The magnetization that
determines the torque is the total magnetization  which, in that scenario, can lie
close to $\phi=0$, though the Ho moments lie only along easy directions locally.
We have found, that in these two new (intermediate and border) phases,
the net magnetization appears to rotate in steps toward the hard axis as the applied field
is rotated toward that direction, suggesting that they may both be states consisting
of mixture of two or more phases in a ratio that changes with applied field.
The discovery of ``frustration'' in the magnetic phases of HoNi$_{2}$B$_2$C for
the field direction along the hard axis $\langle 100 \rangle$ is an important new result
of the present study.
\par
The new phenomena found in this study raises the question: why were they not revealed
in the longitudinal magnetization measurements \cite{canf}? Obvious reasons may be
insensitivity to the component of $\vec{M}$ normal to the field and
the rather large error bars for orientation of the field in the $ab$-plane of
HoNi$_{2}$B$_2$C crystal lattice, which in Ref. \cite{canf} were about $\pm 4^{\circ}$. In
the present study this error is far less ($\pm 1^{\circ}$), but the main advantages
of torque magnetometry are that the relative angular accuracy (when rotating
a sample consistently in only one direction) is very high and that it is very
sensitive to the normal component of $\vec{M}$ which is subject to fluctuations
near $\phi=0$. The data obtained show quite consistent, not random changes in the torque
magnitude for angle variations about 0.25$^{\circ}$ or even less. As a result,
the angular torque dependences consist of hundreds of points that allow detection of
important features of the torque angular behavior.

\section{Conclusions}
These torque measurements provide not only a check of the
known results and refinement of the angular diagram for
metamagnetic transitions in HoNi$_{2}$B$_2$C, but also clarify some important features
of these metamagnetic states and, moreover, indicate
new states. Indications of a magnetically inhomogeneous state of the
AFM phase at low magnetic field, which is likely determined by a combined influence
of multidomain magnetic structure and flux trapping in the superconducting state were found.
The first precise determination of the magnetization direction in the
($\uparrow\uparrow\rightarrow$) phase was made. This direction varies with  the magnitude
of the applied field which does not agree with model predictions \cite{kalats,amici}.
Two new phase boundaries parallel to the hard axis in the $ab$-plane polar field plot
(Fig. 1) were indicated in a region previously ascribed to the non-collinear
($a^{*}$ or $b^{*}$ modulated) phase. ``Frustration'' was observed when the
direction of magnetic field is close to the hard axis $\langle 100
\rangle$ ($-6^{\circ} \lesssim\theta \lesssim 6^{\circ}$) and
tentatively explained by a mixed (two-domain) state of the system with moments
aligned along different equivalent easy axes.
These torque magnetometry measurements indicate a more complicated magnetic behavior and
suggest important parts of the phase diagram where more detailed neutron scattering
measurements should be focussed.

\acknowledgments
The authors acknowledge P. C. Canfield for providing single crystal samples and
V. Pokrovsky and I. Lyuksyutov for helpful comments.
The work reported here was supported in part by the Robert A.
Welch Foundation (Grant No. A-0514), the Telecommunications and
Information Task Force at Texas A\&M University, and
the National Science Foundation (Grant Nos. DMR-0315476, and
DMR-0422949). Partial support of the U. S. Civilian Research and
Development Foundation for the Independent States of the Former
Soviet Union (Project No. UP1-2566-KH-03) is acknowledged.




\end{document}